 \documentclass[aps,pra,superscriptaddress,amsmath,amssymb,preprintnumbers,twocolumn,floatfix,showpacs,showkeys,10pt]{revtex4-1}
 \usepackage{amssymb} \usepackage{epsfig}
 \begin{document}
  \title{Generation and protection of steady-state quantum correlations due to quantum channels with memory  }

\author{You-neng Guo}
\email{guoxuyan2007@163.com}
\affiliation{ Department of Electronic and Communication Engineering, Changsha University, Changsha, Hunan
410022, People's Republic of China}
\affiliation{Hunan Province Key Laboratory of Applied Environmental Photocatalysis, Changsha University, Changsha, Hunan
410022, People's Republic of China}
\author{Mao-fa Fang}
\email{mffang@hunnu.edu.cn}
\affiliation{ Key Laboratory of Low-Dimensional Quantum Structures and
Quantum Control of Ministry of Education, and Department of Physics,
Hunan Normal University, Changsha 410081, People's Republic of
China}
\author{Guo-you Wang}
\email{gywang04@163.com}
\affiliation{College of Science, Hunan University of Technology, Zhuzhou 412008, People's Republic of China}
\author{Ke Zeng}
\email{zk92@126.com}
\affiliation{ Department of Electronic and Communication Engineering, Changsha University, Changsha, Hunan
410022, People's Republic of China}
\affiliation{Hunan Province Key Laboratory of Applied Environmental Photocatalysis, Changsha University, Changsha, Hunan
410022, People's Republic of China}

\begin{abstract}
We have proposed a scheme of the generation and preservation of two-qubit steady-state quantum correlations through quantum channels where successive uses of
the channels are correlated. Different types of noisy channels with memory, such as amplitude damping, phase-damping, and depolarizing channels have been taken into account. Some analytical or numerical results are presented. The effect of channels with memory on dynamics of quantum correlations has been discussed in detail. The results show that, steady-state entanglement between two initial qubits without entanglement subject to amplitude damping channel with memory can be generated. The entanglement creation is related to the memory coefficient of channel $\mu$. The stronger the memory coefficient of channel $ \mu$ is, the more the entanglement creation is, and the earlier the separable state becomes the entangled state. The result also shows that there exists nonlocality in the absence of entanglement. Besides, we compare the dynamics of entanglement with that of quantum discord when a two-qubit system is initially prepared in an entangled state. We show that entanglement dynamics suddenly disappears, while quantum discord displays only in the asymptotic limit. Furthermore, two-qubit quantum correlations can be preserved at a long time in the limit of $\mu\rightarrow1$.

 \end{abstract}

 \maketitle
\section{Introduction}
Quantum correlations, such as quantum entanglement and quantum discord, have been proposed as the crucial resources applications in quantum information processing \cite{1}. Entanglement is a special type of quantum correlations, and it has been proven to be very important for quantum communication networks and quantum computations \cite{2,2a}. On the other hand, quantum discord which captures more general quantum correlations than entanglement, has initially been introduced by Olliver and Zurek \cite{3}. It has been proven that quantum discord can exist without entanglement and it provides quantum advantages for some quantum information tasks \cite{4,4a}. Over the past decade, the study of quantum correlations applying to quantum communication tasks and quantum computational models has attracted much attention,
they however, are usually very difficult to be created, maintained, and manipulated in realistic systems.
Furthermore, quantum correlations are fragile and prone to environmental effect due to any systems inevitably interacting with their external environments which result in quantum systems' decoherence as well as disentanglement, therefore it is very important and necessary to create and preserve the quantum correlations in the field of quantum information.

The generation and preservation of bipartite or multipartite quantum correlations have been widely studied both theoretically \cite{5,6,7,8,9,10,11,12,13,14,15,15a,15b,15c} and experimentally \cite{16a,16b,16,17,18,19} in recent years. Usually, people use of methods to create entanglement either indirect coupled quantum systems or direct coupled quantum systems \cite{20,21,22,23,24,25,26}. The former is to introduce a simple ancillary system, whose interactions with the decoupled subsystems lead to their indirect interactions with each other. For instance,
the entanglement between two qubits can also be created by immersing in a common heat bath environment \cite{21}. Recently, it happens that even a dissipative common environment or thermal electromagnetic field is able to induce entanglement among subsystems \cite{22,24}.

At the same time, there are many methods developed to protect quantum correlations from decoherence, such as error-correcting codes \cite{27,28}, strategies based on decoherence-free subspaces \cite{29,30}, or using detuning modulation \cite{31}. Besides, dynamical decoupling \cite{32,33,33a} as well as quantum Zeno effect \cite{34}can also be used to protect bipartite quantum correlations by tackling decoherence. As mentioned above these previous studies, the back-action of the environment and the memory effect of the environment play a significant role in combatting decoherence. Recently, using the weak measurement and its reversal measurement to protect quantum correlations from the amplitude damping decoherence has been demonstrated \cite{35,36,37,38,39,40}. In addition, there have been several interesting works \cite{41,42} that focused
on the protection of the long time limit of quantum correlations through the addition of qubits.

In this paper, we aim to study how the correlated noise acting on consecutive uses of channel influences
the quantum correlations dynamics. In real physical quantum transmission channels, the noise over consecutive uses
provide a natural theoretical framework for the study of any noisy quantum communication since they were first considered by C. Macchiavello and G.M. Palma \cite{43} to study the problem of classical capacity. Later, the study of information transmission over a correlated channel with an arbitrary degree of memory in quantum information process has attracted much attention \cite{43a,43b,44,45,46,47}. Recently, a report about the effect of correlated noise on the entanglement of X-type state of the Dirac fields in the non-inertial frame has been investigated by M. Ramzan \cite{44}. We here, have analyzed the generation and preservation of two-qubit quantum correlations through quantum channels where successive uses of the channels are correlated. Different types of noisy channels with memory, such as amplitude damping, phase-damping, and depolarizing channels have been taken into account. The effect of channels with memory on dynamics of quantum correlations has been discussed in detail. The results show that, steady-state entanglement between two independent qubits initially without entanglement subject to amplitude damping channel with memory can be generated. We observe that there exists nonlocality quantified by quantum discord in the absence of entanglement. Besides, we compare the dynamics of entanglement with that of quantum discord when a two-qubit system is initially prepared in an entangled state. We show that entanglement dynamics suddenly disappears,
whlie quantum discord vanishes only in the asymptotic limit. Two-qubit quantum correlations can be preserved at a long time due to the channels with prefect memory.

The layout is as follows: In Sec. \textrm{II}, we illustrate the initial states and noise channels. In  Sec. \textrm{III}, we devote to examining generation and protection of two-qubit quantum correlations in different types of noisy channels with memory. Finally, we give the conclusion  in Sec. \textrm{IV}.

\section{Initial states and noise model}  
We begin with a brief description of quantum-memory
channels. As we all know, there are two different types of quantum channels including memoryless channels and memory channels. The simplest models for quantum channels are memoryless, when environmental correlation time is smaller than the time between consecutive
uses, so that at each channel use the environment back action can be negligible. Namely, the system undertakes the same quantum channel $\varepsilon$, in which independent noise acts on each use. Suppose $N$ times uses of this channel, then we have $\varepsilon_{N}=\varepsilon_{}^{\otimes N}$.
However, real systems among subsequent channel uses exhibit some correlations, this can happen when environmental correlation time is longer than the time between consecutive uses, so that the channel acts dependently on each channel input, $\varepsilon_{N} \neq \varepsilon_{}^{\otimes N}$. These kinds of channels are called memory channels.

In what follows we consider $N$ channel uses.
Given an input state $\rho$, a quantum channel $\varepsilon$ is defined as a
completely positive, trace-preserving map from input-state
density matrices to output-state density matrices,
\begin{equation}
\varepsilon(\rho)=\sum_{i}E_{i}\rho E_{i}^{\dagger}
\end{equation}
where $E_{i}=\sqrt{P_{i_{1}...i_{N}}}A_{i}$ are the Kraus operators of the channel which satisfy the completeness relationship, and $\sum_{i}P_{i_{1}...i_{N}}=1$. Here $P_{i_{1}...i_{N}}$
can be interpreted as the probability that a
random sequence of operations is applied to the sequence of
$N$ qubits transmitted through the channel. For a memoryless
channel, these operations $A_{i}$ are independent, and we have $P_{i_{1}...i_{N}}=P_{i_{1}}P_{i_{2}}...P_{i_{N}}$. However, for a channel with memory, these operations are time-correlated, $P_{i_{1}...i_{N}}=P_{i_{1}}P_{i_{2}|i_{1}}...P_{i_{N}|i_{N-1}}$, here $P_{i_{N}|i_{N-1}}$ is the conditional probability for that operation. For simplicity,  we consider the Kraus
operators for two consecutive uses of a channel with
partial memory are
\begin{equation}
E_{i,j}=\sqrt{P_{i}[(1-\mu)P_{j}+\mu\delta_{i,j}]}A_{i}\otimes A_{j}
\end{equation}
where $0\leq \mu \leq 1$ is the memory coefficient of channel.

Following, we will focus on the noisy channels (e.g. amplitude damping, phase-damping, and depolarizing channels) with time-correlated Markov noise for two consecutive uses. Based on the Kraus operator approach, for any initial state $\rho$, the finial state under noise is given by \cite{43,45}
\begin{equation}
\varepsilon(\rho)=(1-\mu)\sum_{i,j}E_{i,j}\rho E_{i,j}^{\dagger}+\mu\sum_{k}E_{k,k}\rho E_{k,k}^{\dagger}
\end{equation}
One can understand the above expression that the same operation is
applied to both qubits with probability $\mu$ while with probability $1-\mu$ both operations are uncorrelated.

\begin{figure}[htpb]
  \begin{center}
    \includegraphics[width=7.5cm]{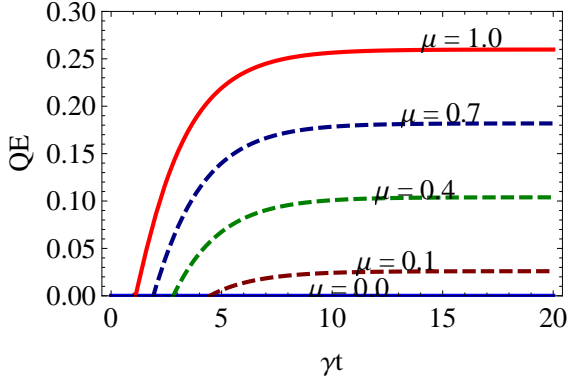}
 \caption{\label{Fig1}(Color online) Quantum entanglement of a two-qubit system initially prepared in an unentangled state under amplitude damping channels with memory is plotted against $\gamma t$ for different $\mu$. Other parameters: $\alpha=0.5,r=0.3$.}
\end{center}
\end{figure}

\begin{figure}[htpb]
  \begin{center}
    \includegraphics[width=7.5cm]{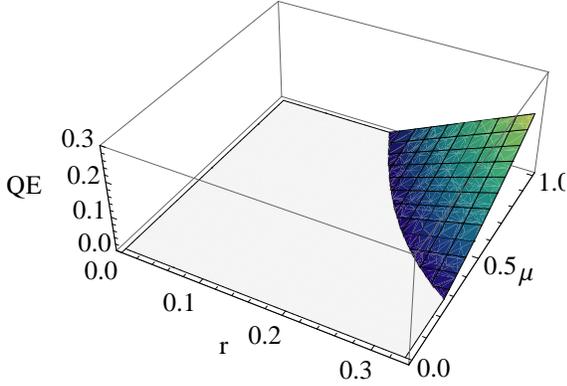}
 \caption{\label{Fig2}(Color online) Generation of quantum entanglement is plotted against $r$ and  $\mu$ under amplitude damping channels with memory. Other parameters: $\alpha=0.5,p=0.95$.}
\end{center}
\end{figure}

\begin{figure}[htpb]
  \begin{center}
   \includegraphics[width=7.5cm]{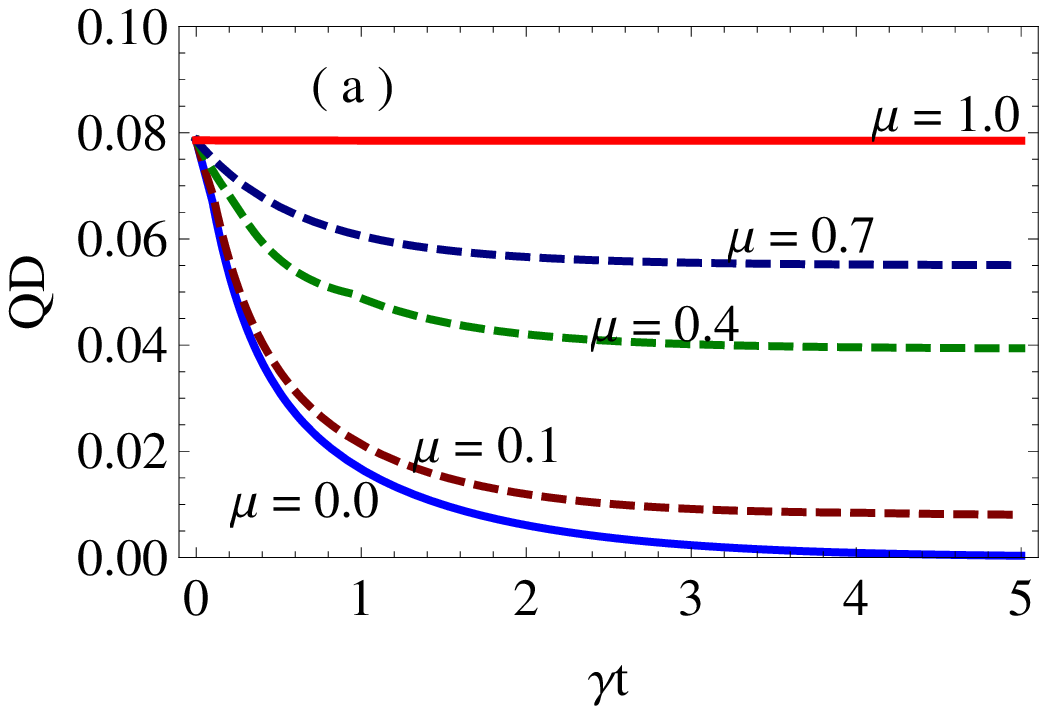}
   \includegraphics[width=7.5cm]{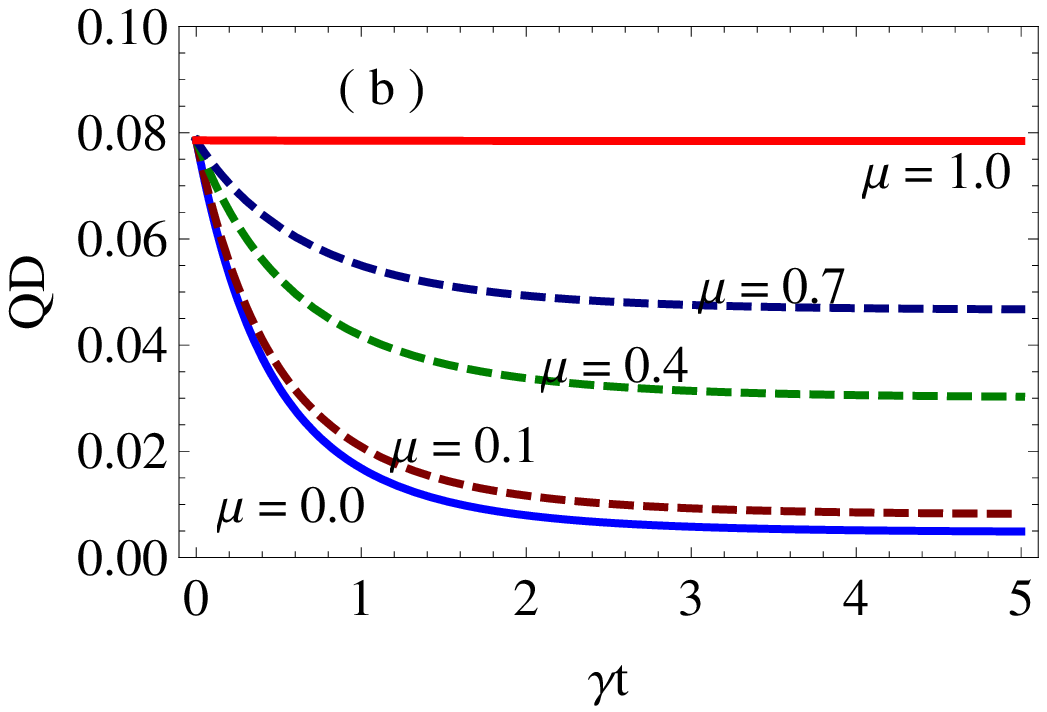}
   \includegraphics[width=7.5cm]{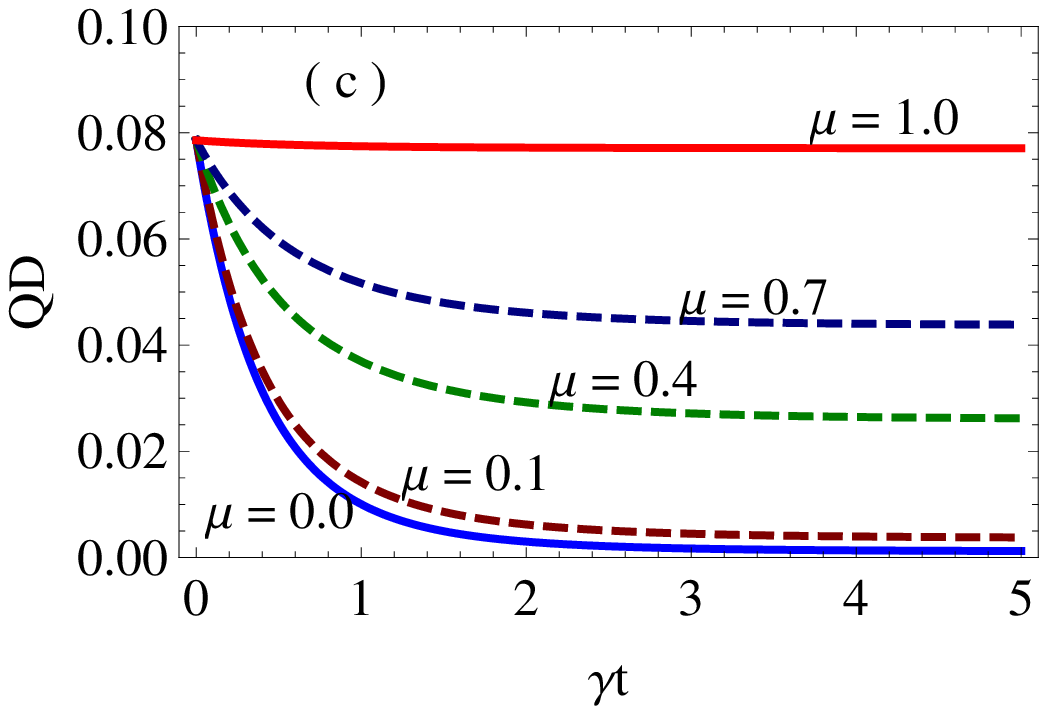}
 \caption{\label{Fig3}(Color online) Quantum discords of a two-qubit system initially prepared in an unentangled state under different quantum channels with memory are plotted against $\gamma t$ for different $\mu$. (a) Amplitude damping channels, (b) phase damping channels, (c) depolarizing damping channels. Other parameters: $\alpha=0.5,r=0.3$.}
\end{center}
\end{figure}

\begin{figure}[htpb]
  \begin{center}
   \includegraphics[width=7.5cm]{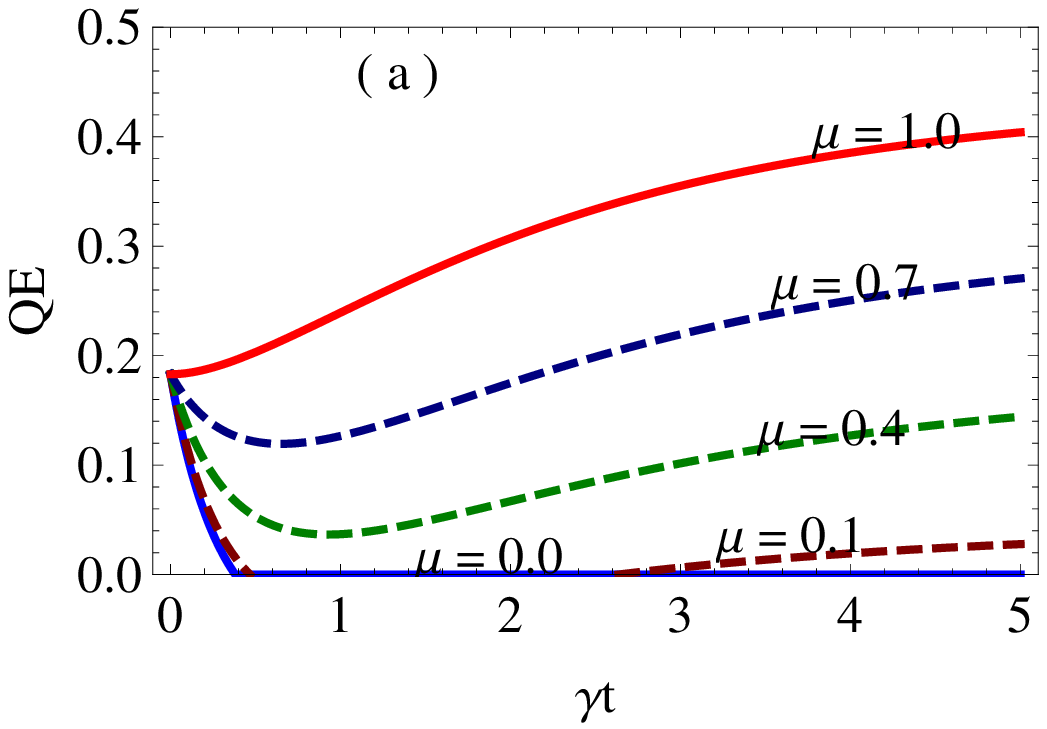}
   \includegraphics[width=7.5cm]{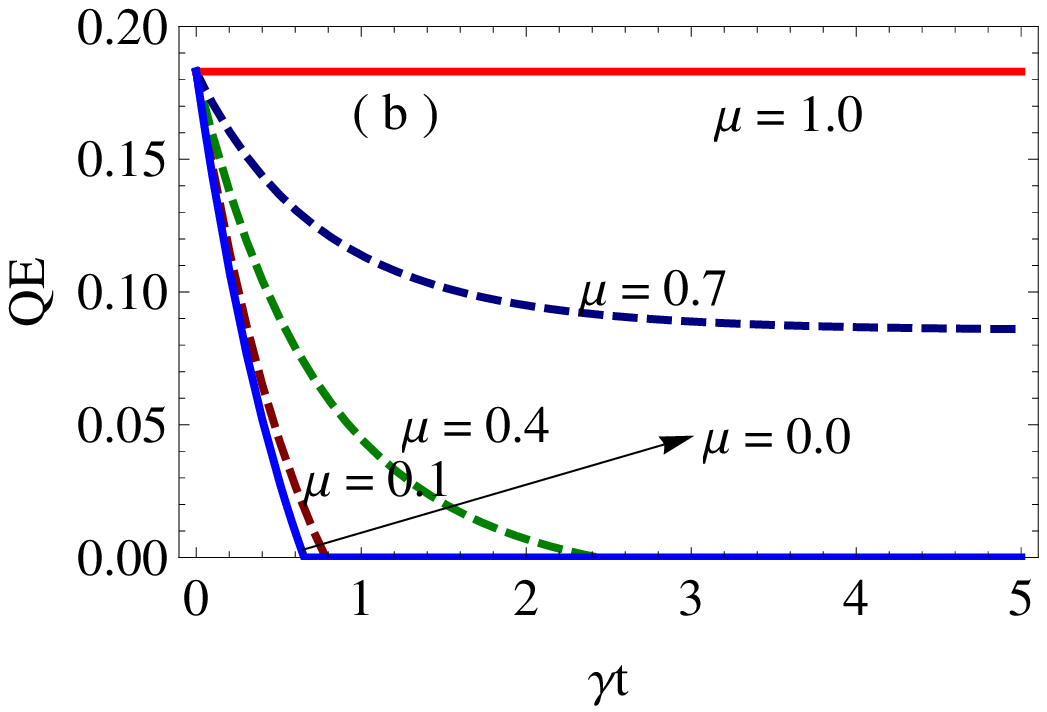}
   \includegraphics[width=7.5cm]{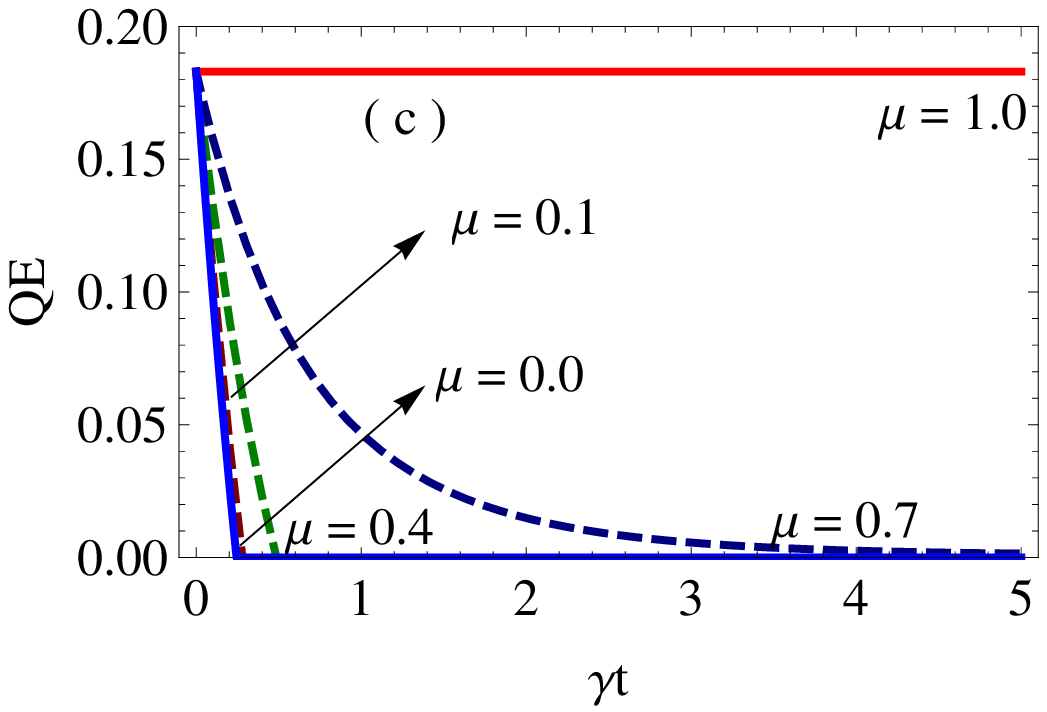}
 \caption{\label{Fig4}(Color online)Quantum entanglements of a two-qubit system initially prepared in an entangled state under different quantum channels with memory are plotted against $\gamma t$ for different $\mu$. (a) Amplitude damping channels, (b) phase damping channels, (c) depolarizing damping channels. Other parameters: $\alpha=0.5,r=0.5$.}
\end{center}
\end{figure}

\subsection{Amplitude damping channel with memory}
Amplitude damping channel which is used to characterize spontaneous emission describes the energy dissipation from a quantum system. The Kraus operators for a single qubit are given by
\begin{equation}A_{0}= \left(
\begin{array}{ c c c c l r }
\sqrt{1-p} & 0  \\
0 & 1  \\
\end{array}
\right)
\end{equation}
\begin{equation}A_{1}= \left(
\begin{array}{ c c c c l r }
0 & 0  \\
\sqrt{p} & 0  \\
\end{array}
\right)
\end{equation}
where $p\equiv1-\exp(-\gamma t)$ is the single qubit damping rate. Suppose that two qubits pass through the quantum dephasing channel, the environmental correlation time is smaller than the time between consecutive uses, and then the quantum amplitude damping channel with uncorrelated noise can similarly be defined as the following Kraus operators
\begin{equation}
E_{i,j}=A_{i}\otimes A_{j},(i,j=0,1)
\end{equation}
Based on previous analysis \cite{43,45}, an amplitude-damping channel with finite memory for two consecutive uses, the task of constructing Kraus operators $E_{k,k}$
for the amplitude-damping channel is given by
\begin{equation}E_{00}= \left(
\begin{array}{ c c c c l r }
\sqrt{1-p} & 0 & 0 & 0  \\
0 & 1 & 0 & 0  \\
0& 0 & 1& 0 \\
0& 0 & 0 & 1\\
\end{array}
\right)
\end{equation}
\begin{equation}E_{11}= \left(
\begin{array}{ c c c c l r }
0 & 0 & 0 & 0  \\
0 & 0 & 0 & 0 \\
0 & 0 & 0 & 0 \\
\sqrt{p} & 0 & 0 & 0  \\
\end{array}
\right)
\end{equation}
Consider the following initial states
\begin{equation}
\rho(0)=r|\Psi\rangle \langle\Psi|+\frac{1-r}{4}I
\end{equation}
where $\Psi=\sqrt{1-\alpha^2}|01\rangle+\alpha|10\rangle$  corresponds to the Bell-like states with $0\leq \alpha\leq 1$, and $r$ indicates the purity of the initial states.
\begin{figure}[htpb]
  \begin{center}
   \includegraphics[width=7.5cm]{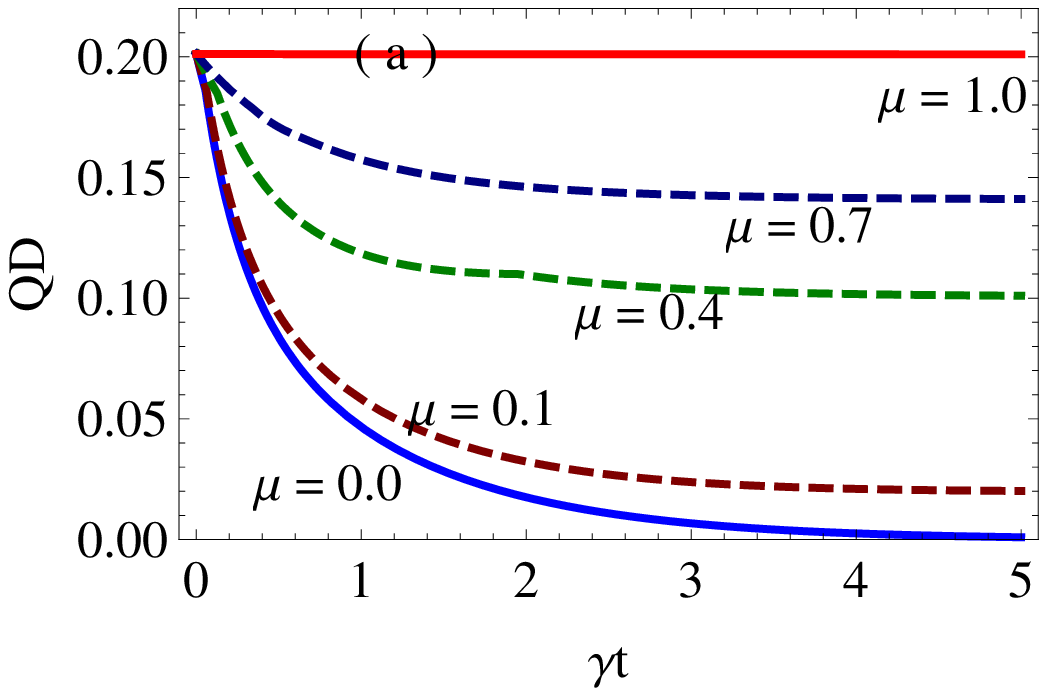}
   \includegraphics[width=7.5cm]{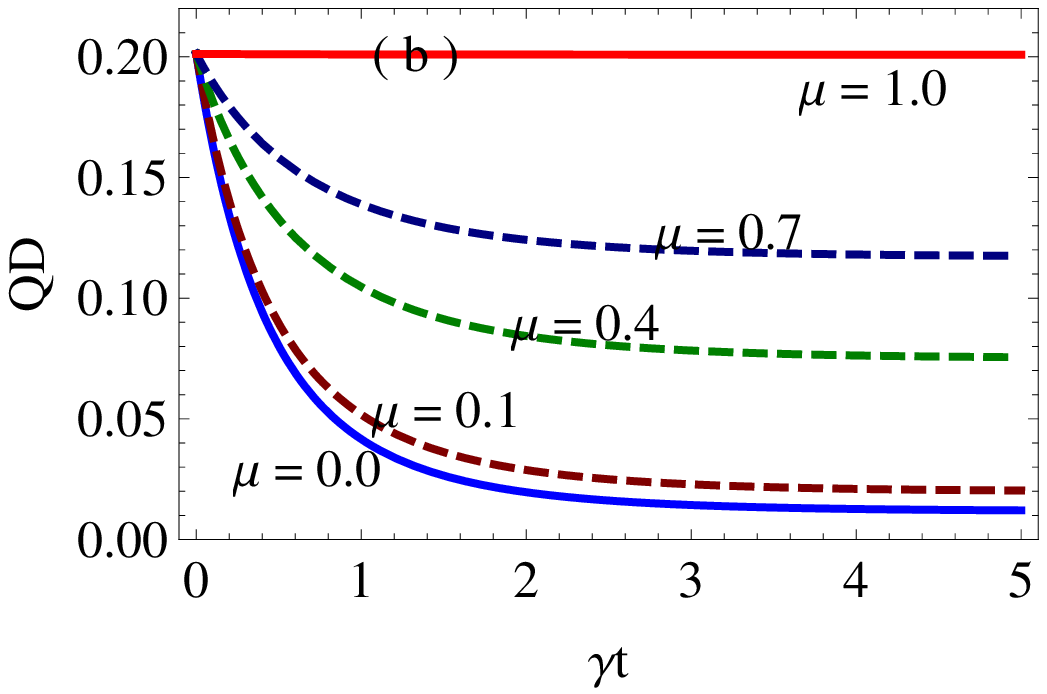}
   \includegraphics[width=7.5cm]{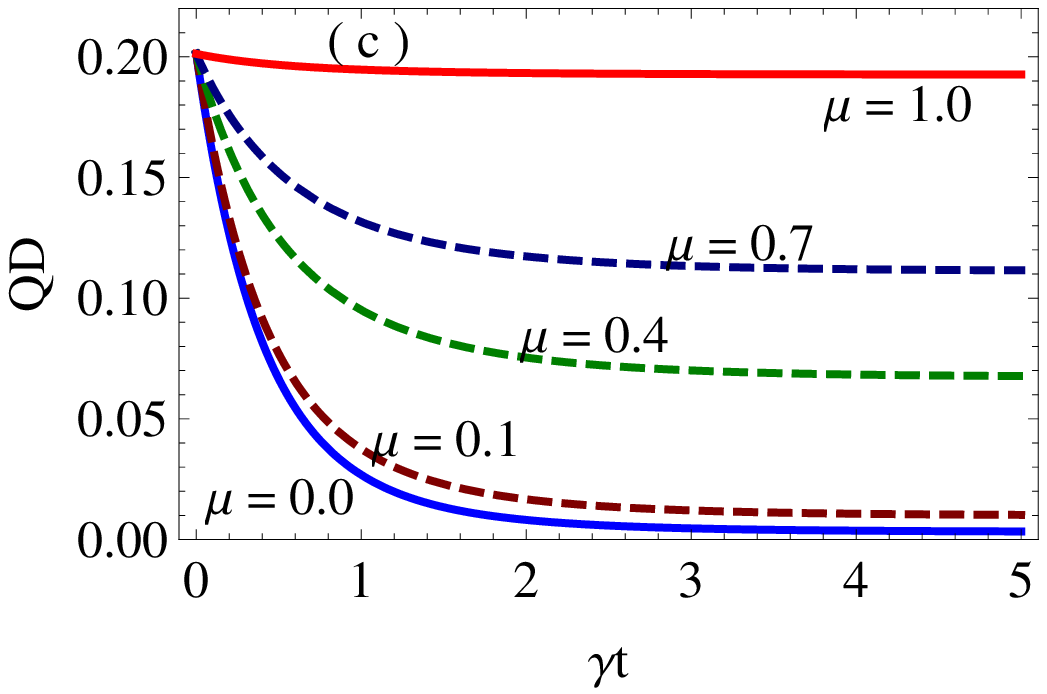}
 \caption{\label{Fig5}(Color online)Quantum discords of a two-qubit system initially prepared in an entangled state under different quantum channels with memory are plotted against $\gamma t$ for different $\mu$. (a) Amplitude damping channels, (b) phase damping channels, (c) depolarizing damping channels. Other parameters: $\alpha=0.5,r=0.5$.}
\end{center}
\end{figure}
According to Eq. (3), through calculations, it is not difficult to get the density matrix of a two qubits system under an amplitude-damping channel with time-correlated Markov noise for two consecutive uses,

\begin{equation}\rho_{AB}(t)= \left(
\begin{array}{ c c c c l r }
\rho_{11}(t) & 0 & 0 & 0 \\
0 & \rho_{22}(t) & \rho_{23}(t) & 0 \\
0 & \rho_{32}(t) & \rho_{33}(t) & 0 \\
0 & 0 & 0 & \rho_{44}(t) \\
\end{array}
\right)
\end{equation}
whose the density matrix elements are
\begin{equation}
\rho_{11}=\frac{1}{4}(1- p) (1 -r) [1 - p (1 - \mu)], \nonumber
\end{equation}
\begin{equation}
\rho_{22}=\frac{1}{4}[1+r(3-4\alpha^2)-4p r(1-\alpha^2)(1-\mu)-p^2(1-r)(1-\mu)], \nonumber
\end{equation}
\begin{equation}
\rho_{33}=\frac{1}{4}[1-r(1-4\alpha^2)-p^2(1-r)(1-\mu)-4p r \alpha^2 (1-\mu)],\nonumber
\end{equation}
\begin{equation}
\rho_{44}=\frac{1}{4}[1 -r+p^2(1-r)(1-\mu)+2p-p\mu+p r(2-3\mu)], \nonumber
\end{equation}
\begin{equation}
\rho_{23}=\rho_{32}^{\ast}=r \alpha\sqrt{1-\alpha^2} [1-p (1-\mu)].
\end{equation}

\subsection{Phase-damping channel with memory}
Phase-damping channel describes a quantum noise with loss of quantum phase information without loss of energy. The Kraus operators for a single qubit are defined in terms of the Pauli operators $\sigma_{0}=I$ and $\sigma_{3}$.

Suppose that two qubits pass through the quantum dephasing channel,  the channel with uncorrelated noise can similarly be defined by the following Kraus operators
\begin{equation}
E_{i,j}=\sqrt{P_{i}P_{j}}\sigma_{i}\otimes \sigma_{j}
\end{equation}

and two consecutive uses of the dephasing channel with partial memory, the Kraus operators $E_{k,k}$ is given as \cite{46}
\begin{equation}
E_{k,k}=\sqrt{P_{k}}\sigma_{k}\otimes \sigma_{k}
\end{equation}
where $i,j,k=(0,3)$, and $P_{0}=1-p$, $P_{3}=p$, where $p\equiv\frac{1}{2}[1-\exp(-\gamma t)]$.
Similarly according to Eq. (3), the density matrix elements of a two qubits system under a phase-damping channel with time-correlated Markov noise for two consecutive uses are
\begin{equation}
\rho_{11}=\rho_{44}=\frac{1}{4}(1-r), \nonumber
\end{equation}
\begin{equation}
\rho_{22}=\frac{1}{4}[1+r(3-4\alpha^2)], \nonumber
\end{equation}
\begin{equation}
\rho_{33}=\frac{1}{4}[1-r(1-4\alpha^2)],\nonumber
\end{equation}
\begin{equation}
\rho_{23}=\rho_{32}^{\ast}=r \alpha\sqrt{1-\alpha^2} [(1-p)^2 (1-\mu)+\mu].
\end{equation}

\subsection{Depolarizing channel with memory}
Depolarizing  channel is another important type of quantum noise, it describes the process in which the density matrix is dynamically replaced by the state $I/2$. $I$ denoting identity matrix of a qubit. The Kraus operators for a single qubit are given by
\begin{equation}
A_{i}=\sqrt{P_{i}}\sigma_{i},(i=0,1,2,3)
\end{equation}
where $P_{0}=1-p$, $P_{1}=P_{2}=P_{3}= p/3$, and $p\equiv\frac{1}{2}[1-\exp(-\gamma t)]$.

Assume that the environmental correlation time is smaller than the time between consecutive
uses, a channel is memoryless and corresponds to its Kraus operators
\begin{equation}
E_{i,j}=\sqrt{P_{i}P_{j}}\sigma_{i}\otimes \sigma_{j},(i,j=0,1,2,3)
\end{equation}
Here we will consider the case
of two consecutive uses of a channel with partial memory,
the task of constructing Kraus operators $E_{k,k}$
for the non-Pauli depolarizing channel is given \cite{47}
\begin{equation}
E_{k,k}=\sqrt{P_{k}}\sigma_{k}\otimes \sigma_{k},(k=0,1,2,3)
\end{equation}

According to Eq. (3), the density matrix elements of a two qubits system under a depolarizing channel with time-correlated Markov noise for two consecutive uses
are
\begin{equation}
\rho_{11}=\rho_{44}=\frac{1}{36}[9 + r(36\alpha^2-9+16p^2(1-\mu) - 24 p (2 \alpha^2-\mu))], \nonumber
\end{equation}
\begin{equation}
\rho_{22}=\frac{1}{36}[9 +r (27-36\alpha^2+16p^2(1-\mu)+24p(2\alpha^2+\mu-2))], \nonumber
\end{equation}
\begin{equation}
\rho_{33}=\frac{1}{36}[9+r(36\alpha^2-9+16p^2(1-\mu)+24p(\mu-2\alpha^2))],\nonumber
\end{equation}
\begin{equation}
\rho_{23}=\rho_{32}^{\ast}=\frac{1}{9}r \alpha\sqrt{1-\alpha^2}[(3-4p)^2(1-\mu)+9\mu].
\end{equation}

\section{Generation and protection of two-qubit quantum correlations under quantum channels with memory }  
Before investigating the generation and protection of two-qubit quantum correlations under quantum channels with memory, let us review the correlation measures quantified by quantum discord (QD) \cite{3} which is defined by
subtracting the classical correlation $C(\rho_{AB})$ from the total amount of correlation $I(\rho_{A:B})$, namely, $QD(\rho_{AB})=I(\rho_{A:B})-C(\rho_{AB})$. The total correlation is quantified by the quantum mutual information $I(\rho_{A:B})=S(\rho_{A})+S(\rho_{B})+S(\rho_{AB})$, where $\rho_{A}$($\rho_{B}$) is the reduced matrix of $\rho_{AB}$ by tracing out $B$($A$). The classical correlation $C(\rho_{AB})$ is defined as $C(\rho_{AB})=S(\rho_{A})-\min_{\{\prod_{k}^{B}\}}S(\rho_{A|B})$, where $S(\rho)=-tr(\rho \log_{2}\rho)$ is the von Neumann entropy. Note that the minimum is taken over the set of positive operator valued measurement $\{\prod_{k}^{B}\}$
on subsystem B,  $S(\rho_{A|B})$ is the conditional entropy for the subsystem A.

It is worthy pointing out that the calculation of classical correlation involves a potentially complex optimization process even numerically, usually there is no general analytical expression of discord except for the simplest case of two-qubit state. Such as for a bipartite quantum X-state
described by the density matrix $\rho_{AB}$, an expression of the quantum discord is given \cite{48}
\begin{equation}
\begin{array}{ c c c c l r }
QD(\rho_{AB})=\min(Q_{1},Q_{2})\\
\end{array}
\end{equation}

where

$Q_{i}=H(\rho_{11}+\rho_{33})+\Sigma_{i=1}^{4}\epsilon_{i}\log_{2}\epsilon_{i}+D_{j}$,

$D_{1}=H(\frac{1+\sqrt{[1-2(\rho_{33}+\rho_{44})]^2+4(|\rho_{14}|+|\rho_{23}|)^2}}{2})$,

$D_{2}=-\Sigma_{i}\rho_{ii}\log_{2}\rho_{ii}-H(\rho_{11}+\rho_{33})$,

$H(x)=-x\log_{2}x-(1-x)\log_{2}(1-x)$.

Besides, to investigate the two-qubit quantum correlations dynamics,
we also use concurrence as the quantifier \cite{49}. For the density matrix of a bipartite system has a form given by Eq. (10), the concurrence reads
\begin{equation}
\begin{array}{ c c c c l r }
QE(\rho_{AB})=2\max(0,|\rho_{23}|-\sqrt{\rho_{11}\rho_{44}},|\rho_{14}|-\sqrt{\rho_{22}\rho_{33}})\\
\end{array}
\end{equation}
Evidence suggests entanglement quantified by concurrence varies from $QE=0$ for a separable state to $QE=1$ for a maximally entangled state.

In the following, we adopt both quantum discord and concurrence as a measure of quantum correlations for a two uncoupled qubits system initially prepared in Eq.(9) successive uses of noisy channels with memory, such as amplitude damping, Phase-damping, and depolarizing channels.
One can easily check the initial state given by Eq.(9), the concurrence $QE(\rho_{0})=0$ for the purity $0\leq r\leq1/3$, while the purity $1/3 < r \leq1$ corresponds to the concurrence $QE(\rho_{0})>0$.

Firstly, we consider the case where the initial state given by Eq.(9) is prepared in an unentangled state (e.g.$\alpha=0.5,r=0.3$), when two qubits pass through the quantum channels with memory. Due to the amplitude damping channels with memory, there is no entanglement at earlier times $QE(\rho_{0})=0$, and at some time the quantum entanglement of two-qubit system starts to build up (see Fig. 1). This phenomenon is called delayed sudden birth of entanglement \cite{24}. Note that the degree of entanglement depends on the memory coefficient of channel $\mu$. The stronger the memory coefficient of channel $ \mu$ is, the more the entanglement creation is, and the earlier the separable state becomes the entangled state. This implies the amplitude damping channels with memory can locally create entanglement, and it can be preserved an unentangled state at a steady entangled state in the presence of the quantum channels with memory. Surprisingly, even though Phase-damping, and depolarizing channel with memory, two-qubit system can not be induced entanglement.

To get a better understanding of the effects of the amplitude damping channel with memory on the entanglement creation,
we plot Fig.2 to show the entanglement creation as a function of initial state purity $r$ and memory coefficient of channel $\mu$ for $p=0.95$. One can see that the entanglement creation is not only depended on the memory coefficient of channel $\mu$ but also on the initial state chosen. However, when chosen $p=1$, we find that, for any $\alpha$ and $0\leq r\leq1/3$, $QE(\rho_{t})=r\alpha\sqrt{1-\alpha^2}\mu \geq0$, which means that two uncoupled qubits system initially prepared in any separable states given by Eq.(9) can be induced entanglement with the help of the amplitude damping channels with memory. Our result indicates that the amplitude damping channels with memory can be locally created entanglement, which is attributed to $\mu$ as well as $p$.

Fig.3 shows quantum discord dynamics for two initial unentangled qubits (e.g.$\alpha=0.5,r=0.3$) in different quantum channels with memory.
The results show that in a case where there is initially no entanglement between two-qubit system but there exists quantum discord. This means quantum discord captures more general quantum correlations than entanglements. It is proven that the states with nonzero quantum discord but not entanglement
are responsible for the efficiency of a quantum computer \cite{4,4a}. Besides, quantum correlations quantified by quantum discord exhibit in the asymptotic limit, and can be well protected in the presence of channels with memory. Especially, as the memory coefficient of channel $\mu$ increases, quantum discord can be protected more effectively. In the limit of $\mu\rightarrow1$, the long-living quantum discord preservation can be observed.

Finally, we investigate the dynamics of quantum correlations for two initial entangled qubits (e.g.$\alpha=0.5,r=0.5$) in different types of quantum channels with memory, and compare the dynamics of entanglement with that of quantum discord. The results show that, the entanglement dynamics disappears in a finite time under the influence of noise as shown Fig. 4. This behaviour is named entanglement sudden death, and the results are consistent with Refs. \cite{44}. While quantum discord exhibits only in the asymptotic limit in Fig. 5. This indicates quantum discord is more robustness against the noise than entanglement. Besides, the amount of initial entanglement $QE(0)\approx0.18$ possesses less general quantum correlations than the initial quantum discord $QD(0)\approx0.20$. Take quantum channels with memory into consideration, as we can see that, with the memory coefficient of channel $\mu$ increasing, it can completely circumvent the entanglement sudden death. Particularly, in the limit of $\mu\rightarrow1$, the memory effect can help preserve entanglement and quantum discord at a long time. Moreover, we find, when two qubits pass through amplitude damping channels with memory, the amount of entanglement creation is more than the initial one, as shown Fig.4(a).

\section{Conclusion}  
In conclusion, we have proposed a scheme of the generation and preservation of two-qubit steady state quantum correlations through quantum channels where successive uses of the channels are correlated. Different types of noisy channels with memory, such as amplitude damping, phase-damping, and depolarizing channels have been taken into account. The effect of channels with memory on dynamics of quantum correlations has been discussed in detail.
We conclude the results as follows: Firstly, steady state entanglement between two independent qubits without entanglement subject to amplitude damping channel with memory can be generated, but Phase-damping, and depolarizing channel with memory can not. Secondly, the amplitude damping channels with memory can locally create entanglement, the stronger the memory coefficient of channel $ \mu$ is, the more the entanglement creation is, and the earlier the separable state becomes the entangled state. Thirdly, there is initially no entanglement between two-qubit system but there exists quantum discord in different quantum channels with memory. Finally, we compare the dynamics of entanglement with that of quantum discord when a two-qubit system is prepared in an entangled state. We show that entanglement dynamics suddenly disappears, namely entanglement sudden death occurs, while quantum discord displays only in the asymptotic limit. As the memory coefficient of channel $\mu$ increases, the phenomena of entanglement sudden death can be completely circumvented. Particularly, two-qubit quantum correlations can be preserved at a long time in the limit of
$\mu\rightarrow 1$

\acknowledgments
This work is supported by the National Natural Science Foundation of China (Grant Nos. 11374096).

\label{app:eff-trans}

\end{document}